\documentclass[11pt,preprint2]{aastex}

\newcommand\SNRsmc{1E 0102.2--7219}

             \font\sevenrm=cmr7

\def\mathrm#1{{\hbox{\sevenrm #1}}}

\newcommand\tSNR{t_{\mathrm{snr}}}

\newcommand\EnSN{E_{\mathrm{sn}}}
\newcommand\Mej{M_{\mathrm{ej}}}
\newcommand\rhoej{\rho_{\mathrm{ej}}}

\newcommand\Msun{M_{\odot}}
\newcommand\Vfs{V_{\mathrm{FS}}}

\newcommand\MAZ{M_\mathrm{A0}}
\newcommand\MSZ{M_\mathrm{S0}}

\def\xx#1{\!\times\!10^{#1}}
\newcommand\gameff{\gamma_\mathrm{eff}}

{\catcode`\@=11
\gdef\SchlangeUnter#1#2{\lower2pt\vbox{\baselineskip 0pt\lineskip0pt
\ialign{$\m@th#1\hfil##\hfil$\crcr#2\crcr\sim\crcr}}}}

\def\lesssim{\mathrel{\mathpalette\SchlangeUnter<}}
\newcommand\RT{Rayleigh-Taylor}

\def\kmps{km s$^{-1}$}

\newcommand\CR{\sigma}

\def\pcc{cm$^{-3}$}
\def\muG{$\mu$G}
\def\iec{i.e.,}
\def\egc{e.g.,}

\newcommand\syn{synchrotron}

\newcommand\IC{inverse-Compton}
\newcommand\pion{pion-decay}

\newcommand\alf{Alfv\'en}

\shorttitle{Instabilities in Young Supernova Remnants}
\shortauthors{Blondin \& Ellison}

\begin{document}     
   
\title{Rayleigh-Taylor Instabilities in Young
Supernova Remnants Undergoing Efficient Particle Acceleration}

\author{John M.~Blondin and Donald C.~Ellison}
\affil{Department of Physics, North Carolina State University, Raleigh, NC 27695-8202}
\email{John\_Blondin@ncsu.edu}
\email{Don\_Ellison@ncsu.edu}        

\begin{abstract}
We employ hydrodynamic simulations to study the effects of high shock
compression ratios, as expected for fast shocks with efficient
particle acceleration, on the convective instability of driven waves
in supernova remnants.
We find that the instability itself does not depend significantly on
the compression ratio, $\CR$, with the growth rates and the width of
the {\it mixing} region at saturation being comparable for the range of
ratios we studied; $4 \le \CR \le 21$.
However, because the width of the {\it interaction} region between the
forward and reverse shocks can shrink significantly with increasing
$\CR$, we find that convective instabilities can reach all the way to
the forward shock front if compression ratios are high enough.  Thus,
if supernova blast waves accelerate particles efficiently, we expect
the forward shock to be perturbed with small amplitude, small
wavelength bumps, and to find clumps and filaments of dense ejecta
material in the vicinity of the shock.
In addition and in contrast to situations where $\CR \le 4$, any
enhancement of the radial magnetic field from \RT\ instabilities will
also extend all the way to the shock front and this may help explain
the slight dominance of radial fields long seen in polarization
measurements of young remnants like Tycho.
\end{abstract}        

\keywords{hydrodynamics -- instabilities -- shock waves -- ISM: 
supernova remnants: cosmic rays}
        
\section{INTRODUCTION}

The presence of convective or Rayleigh-Taylor (R-T) instabilities
in young, 
supernova remnants (SNRs) has been argued for a long time
\citep[\egc][]{gul73,shi78,dic89} and examined in detail with
numerical simulations of idealized models
\citep[\egc][]{cbe92,jun96,wan00}.
These instabilities are expected to arise in young, ejecta-dominated SNRs
where a shell of dense, shocked ejecta is gradually decelerated by lower-density
shocked circumstellar material.  This situation is subject to the familiar
R-T instability, which results in fingers of dense ejecta gas protruding
into, and mixing with the shocked circumstellar material.  These instabilities
are relevant to remnants from both Type Ia and Type II explosions, and 
are expected to exist as long as there is a reverse shock present in the
remnant.

These instabilities may be important for mixing ejecta out to, and
possibly past, the forward shock, and for modifying the morphology of
an otherwise spherical outer blast wave and/or reverse shock.  One of
the best studied young SNRs, Cas A exhibits ejecta material out to and
beyond the nominal radius of the forward shock, as evidenced by
optical knots of enriched gas \citep{Fesen96}, as well as spectral
analysis of X-ray knots and filaments \citep{hug00a}.  The outer
blastwave of Cas A, as seen in X-ray emission, is roughly spherical,
but possesses many bumps and protrusions that may be the result of the
convective instability.  
Other examples of irregular blastwaves include the more extreme cases
of SN1986J and SN1993J.
VLBI observations of these SNRs \citep{bar91,BietBR2001} show shells of
radio emission with large radial protrusions, suggesting that very
young SNRs can exhibit large deviations from spherical symmetry.

Another characteristic of young SNRs that may originate with the
convective instability associated with the contact discontinuity is
the slight dominance of a radial magnetic field as deduced from
radio polarization measurements \citep{dic91}.
\citet{bij96} suggested that the polarization observations
could be explained by convective instabilities dragging out an ambient
magnetic field as the  R-T fingers pushed from the
contact discontinuity out toward the forward shock.

Despite the belief that convective instabilities alone should play an
important role in mixing ejecta out to the forward shock and
generating the observed radial magnetic fields, attempts to
demonstrate this have failed.  \citet{cbe92} used a linear
perturbation analysis and two-dimensional hydrodynamic simulations to
study this instability, and found that the mixing region did not reach
the forward shock.  This result was confirmed by \citet{jun96} and
found to hold true even in the presence of rapid cooling of the
shocked ejecta \citep{cbl95}.  \citet{bij96} showed that this limited
mixing region was inconsistent with observations, for it produced a
SNR with a two-shell structure when viewed in radio-synchrotron
emission: an inner shell of strong emission and a dominant radial
magnetic field, and an outer shell of weaker emission with a dominant
tangential field.  \citet{JJ99} extended this work by including a
simplified scheme for injecting and accelerating test-particle
electrons and calculated the radio \syn\ emission self-consistently in
the turbulent fields.  Their work emphasized the role of the reverse shock in
accelerating electrons and producing radio emission.

One possible solution to these problems is the introduction of
small-scale structure into the density of either the 
circumstellar medium (CSM) or the supernova ejecta.
\citet{jjn96} found that including small cloudlets in
the surrounding CSM could enhance the growth of
the Rayleigh-Taylor fingers enough so they reached all the way
to the shock front.  Alternatively, \citet{bbr01} examined the
effects of including low-density bubbles in the supernova ejecta.
They also found an enhanced turbulence in the region between the
forward and reverse shocks, with some clumps of ejecta reaching
out to the forward shock.

All of this previous work concerning R-T instabilities and driven waves
has assumed the supernova ejecta and CSM can
be treated as an ideal gas, typically modeled with an adiabatic index,
$\gamma = 5/3$.\footnote{
For discussions of other types of instabilities in shocks accelerating
cosmic rays see, for example, \citet{DF86} and \citet{MDDD96}.}
However, these assumptions may not be
appropriate for young SNRs, where the strong shock waves are expected
to be efficient accelerators of energetic particles (\iec\ cosmic
rays).
\footnote{
In the heliosphere, where shocks are observed directly with
spacecraft, there is clear evidence that the quasi-parallel earth bow
shock (with a sonic Mach number, $\MSZ < 10$) can place 10-30\% of the
solar wind kinetic energy flux into superthermal particles
\citep[\egc][]{EMP90}. Quasi-perpendicular interplanetary shocks
(IPSs) also accelerate ambient particles but generally with lower
efficiencies due to their lower Mach numbers (generally $\MSZ \lesssim
3$ for IPSs) \citep[\egc][]{Baring97}.  In a few cases, however,
IPSs with higher than average Mach numbers have been observed to
accelerate particles with efficiencies comparable to the
quasi-parallel Earth bow shock \citep{Eichler81,Terasawa99}.  Hybrid
plasma simulations showing direct injection and acceleration of
ambient particles are reasonably consistent with these observations
\citep[\egc][]{STK92}, as are convection-diffusion models
\citep[\egc][]{KJ97}. Roughly, quasi-parallel and
quasi-perpendicular shocks are those with an angle between the
upstream magnetic field and the shock normal of less than $45^\circ$
and greater than $45^\circ$, respectively.
}
It has long been believed that SNRs are the primary sources of
galactic cosmic rays below $\sim 10^{15}$ eV \citep[\egc][]{Axford81}.
However, for this to be the case, the acceleration mechanism must be
efficient and place on the order of $10\%$ or more of the total ejecta
kinetic energy into relativistic particles
\citep[\egc][]{be87,DMV89,Dorfi00}.
Radio observations have long been proof that SNRs produce GeV
electrons and recent evidence of X-ray \syn\ emission in young SNRs
suggests they can produce TeV electrons as well
\citep[\egc][]{Koyama95,MdeJ96,Allen97,Tanimori98,RK99}.  While direct
proof of ion acceleration in SNRs remains elusive, 
indirect evidence exists \citep[e.g.,][]{ESG2001} and there 
is a growing
body of evidence linking the efficient production of cosmic-ray ions
by the forward and reverse shocks with the thermal properties of the
shock-heated X-ray emitting gas 
\citep[\egc][]{DorfiB93,Decourchelle00,Hughes00b}.

One of the important structural effects of efficient particle
acceleration is a shock compression ratio greater than 4
\citep[\egc][]{ee84,BEapj99}.  For an ideal gas in the absence of
particle acceleration, a high Mach number shock produces a compression
ratio,
\begin{equation}
\label{eq:sigma}
\CR \simeq  \frac{\gamma + 1}{\gamma - 1}
\ .
\end{equation}
For the nominal value of $\gamma=5/3$, $\CR = 4$.
However, when efficient particle acceleration occurs, relativistic
particles can be created and substantially contribute to the pressure,
and superthermal particles can escape from the shock. These two
effects combine to produce a shocked plasma which acts to a large
extent like a gas with an effective adiabatic index, $\gameff
< 5/3$, allowing arbitrarily large compression ratios.
In fact, it has been shown \citep{BEapj99} that, 
when injection of particles from the background occurs easily,
nonlinear diffusive
shock acceleration yields
\begin{equation}
\label{eq:machsonic}
\CR \simeq \cases{
 1.3 \, \MSZ^{3/4}
 &if \ \  $1 \ll \MSZ^2 \ll \MAZ$
\ ;
\cr
\noalign{\kern5pt}
 1.5\, M_{A0}^{3/8}
&if \ \ $1 \ll \MAZ < \MSZ^2$
\ .
\cr
}
\end{equation}
Here, $\MSZ$ ($\MAZ$) is the upstream sonic (\alf) Mach number
\citep[see also][]{ke86,Malkov97}.
Furthermore, the increase in compression ratio is accompanied by a
decrease in the post-shock temperature which links the efficient
production of superthermal particles with the thermal properties of
the shock-heated, X-ray emitting gas.

Of particular importance for the work reported here, the increased
compression ratio also implies that the region between the forward and
reverse shocks in young SNRs will be significantly narrower and denser
when acceleration is efficient compared to the case where little
energy is placed in relativistic particles.  The increased density and
narrower spatial extent of the interaction region leads to large
density gradients suggesting that R-T instabilities may have faster
growth rates and larger amplitudes at saturation than is commonly
assumed \citep{Decourchelle00}.  
The dependence of the compression ratio on the sonic and Alfv\'en Mach
numbers implies that these effects will be most pronounced in
SNRs having high shock speeds and relatively low magnetic fields.

In this paper we explore the possibility that the familiar convective
instability in young SNRs may be altered by high compression ratios.
We use a well-tested hydrodynamic simulation of evolving SNRs, only
replacing the standard $\gamma=5/3$ with an effective adiabatic index,
$\gameff < 5/3$.  A softer equation of state (a low
$\gameff$) causes the shocked plasma to be more compressible, yielding
shock compression ratios $\CR > 4$, consistent with those expected
when particle acceleration is efficient.

Somewhat unexpectedly, we find that the convective instability growth
rate and mixing length are almost independent of the compression
ratio. However, since the region between the forward and reverse
shocks narrows as $\CR$ increases, the likelihood that \RT\ fingers
reach and perturb the forward shock is strongly correlated with $\CR$
and, therefore, with the efficiency of cosmic-ray production. 
If $\CR$ is large enough, the convective instabilities do reach and
perturb the forward shock making it likely that clumps and filaments
of dense ejecta material will be found there.

In section \ref{sec:model} we discuss the impact of a low value of
$\gameff$ on a one-dimensional dynamical model of young SNRs.  In
section \ref{sec:hydro} we present hydrodynamic simulations in
one, two, and three dimensions, and in
section \ref{sec:disc} we discuss the approximations and implications
of our numerical results.

\section{A DYNAMICAL MODEL}\label{sec:model}

The dynamical evolution of young SNRs is determined by the interaction
of the stellar ejecta, one to several solar masses of material thrown
out by the SN explosion, with the surrounding circumstellar medium.
In the simplest case, this 
interaction is characterized by a two shock structure: a forward
shock driven ahead of the ejecta into the CSM, and a reverse shock
that decelerates the supersonically expanding ejecta in the vicinity
of the interaction region.  To model this ejecta-dominated evolution,
we begin with the self-similar driven wave (SSDW) model
\citep{rog82a}, although we have found results similar to those
reported here for the exponential ejecta model \citep{dwa98}.

\citet{rog82a} provided spherically symmetric, self-similar solutions
to the structure of ejecta-dominated SNRs under the assumption that
the mass density in both the ejecta and the CSM could be described by
power laws.  
Since the steep power law density profiles for the ejecta
lead to an infinite ejecta mass, we follow Chevalier
and institute a cutoff in
the ejecta density profile at small radii. Therefore, at
radii less than some critical value, $r_c = v_ct$, the ejecta density is
taken to be a spatially constant plateau, while beyond this radius it
follows a power law, i.e.,
\begin{equation}
\rhoej (r,t) = \left\{
\begin{array}{l@{\quad \mbox{for} \quad}l}
\rho_c \, v_c^n \, r^{-n} \, t^{n-3} & r > v_c t \\
\rho_c \, t^{-3} & r < v_c t \\
\end{array} \right.
\end{equation}
where $t$ is time, the constant $\rho_c$ is given by
\begin{equation}
\rho_c = \frac{5n-25}{2\pi n} \EnSN v_c^{-5}
\ ,
\end{equation}
the velocity at the intersection of the ejecta density plateau and the
power law is given by
\begin{equation}
v_c = \left(\frac{10n-50}{3n-9}\frac{\EnSN}{\Mej}\right)^{1/2}
\ ,
\end{equation}
and $n$ is a constant index.
These quantities, $\rho_c$ and $v_c$, are related to the constant $g$ in
\citet{rog82a} by $\rho_c \, v_c^n = g^n$.
Here, $\EnSN$ and $\Mej$ are the kinetic energy and mass of the
ejecta, respectively, and it's clear that fixing the plateau allows us
to relate the self-similar parameters to physical quantities normally
associated with SNRs.
Furthermore, we restrict our examples to times such that $r_c$ is less
than the radius of the reverse shock to avoid complications from
multiply reflected shocks etc., and to make direct comparisons between
our numerical results and self-similar solutions straightforward.

The density of the ambient circumstellar medium  is
also described by a power law in radius in the \citet{rog82a,rog82b}
solution, i.e.,
\begin{equation}
\label{eq:windden}
\rho_a = q \, r^{-s}
\ ,
\end{equation}
where $s=0$ corresponds to a uniform density (in which case $q$ is the
ambient density)
and $s=2$ represents the
density profile of a steady-state stellar wind from the supernova
progenitor of speed, $v_w$, and mass loss rate, $dM/dt$, such that
$q = (dM/dt)/(4 \pi \, v_w)$.

Thus, provided that the edge of the ejecta plateau region has not yet
reached the reverse shock, the radius and expansion velocity of the
forward shock are:
\begin{eqnarray}
\label{eq:FSrad}
R_1 &=& b
\left(\frac{\rho_c v_c^n}{q}\right)^{1/(n-s)}
t^{(n-3)/(n-s)} 
\ , \\
V_1 &=& \left(\frac{n-3}{n-s}\right) \frac{R_1}{t} 
\ ,
\end{eqnarray}
and the fluid variables in the shocked region immediately behind
the forward shock are:
\begin{eqnarray}
\label{eq:FSrad2}
\rho_1 &=& \frac{\gameff + 1}{\gameff - 1} qR_1^{-s}
\ , \\
u_1 &=& \frac{2}{\gameff+1}\left(\frac{n-3}{n-s}\right)\frac{R_1}{t} 
\ , \\
p_1 &=& \frac{2q}{\gameff + 1}\left(\frac{n-3}{n-s}\right)^2 
\frac{R_1^{2-s}}{t^2}
\ ,
\end{eqnarray}
where $\rho$ is the density, $u$ is the fluid speed, $p$ is the pressure, 
and the subscript $1$ indicates 
the forward shock.
Note that all of the dependence on
the adiabatic index in eq.~(\ref{eq:FSrad}) is contained in
the constant $b$, which is related to parameters defined in \citet{rog82a}
through the expression $b = (R_1/R_c) A^{1/(n-s)}$.  The 
calculated values of the constant $A$ and the ratio of the forward
shock radius to the radius of the contact discontinuity for $\gamma=5/3$ for
various $n$ and $s$ are given in Table~1 of \cite{rog82a}.

We note that in an attempt to predict the $\gamma$-ray flux from
\pion, \citet{rog83} considered two-fluid, self-similar solutions
where one fluid was a thermal gas with $\gamma =5/3$ and the other
was a relativistic gas (i.e., cosmic rays) with $\gamma=4/3$.  The
effective adiabatic index was determined by
\begin{equation}
\gameff =
\frac{5 + 3 w}{3(1 + w)}
\ ,
\end{equation}
where $w$ was the (constant) fraction of total pressure made up by relativistic
particles at the shock.  Chevalier did not 
address the possible effects $\gameff$ would have on R-T instabilities nor 
attempt to model the effects of particle escape
during acceleration and assumed $4/3 \le \gameff \le 5/3$ with a
maximum $\sigma =7$ in his calculations.  Chevalier did note, however,
that cosmic rays diffusing from the shocked gas might produce a dense
shell with $\sigma > 7$ similar to that which occurs in the later
radiative phase.

\begin{figure}[!hbtp]              
\plotone{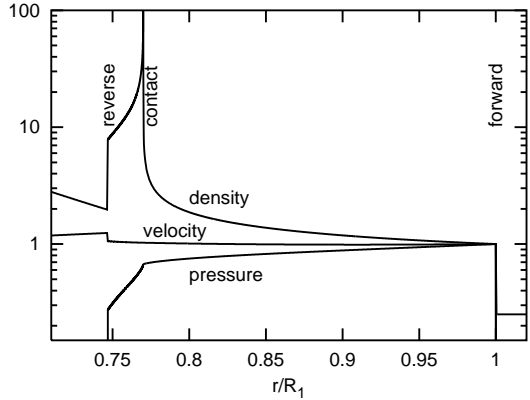}
\figcaption{Internal structure of a SSDW with $n=7$, $s=2$, and
$\gamma = 5/3$ propagating to the right.  The contact discontinuity
separates the shocked ejecta on the left and the shocked CSM on the
right.  The radius is scaled to the position of the forward shock and
the density, pressure, and velocity are scaled to their values
just inside the forward shock.
\label{fig:ssdw}}
\end{figure}

The structure of a SSDW is illustrated in Figure \ref{fig:ssdw},
which shows the radial profiles of density, pressure, and velocity for
values of $n=7$ and $s=2$. The interaction region is bounded by the
forward and reverse shocks, which for this example have compression
ratios $\CR = 4$ corresponding to a high Mach number adiabatic shock
in an ideal fluid with $\gamma = 5/3$.  The pressure is monotonically
increasing from the reverse shock to the forward shock, as all of the
shocked gas is gradually decelerating.

An important characteristic of
these solutions for $s=2$ is the convectively unstable region ahead of 
the contact discontinuity \citep{cbe92}.  This region is marked by opposing
signs of the pressure and density gradients.  The positive
pressure gradient is gradually decelerating the shocked ejecta
and shocked CSM, but the negative density gradient implies that
the denser gas is being decelerated by lower density gas; a situation
that is subject to the familiar R-T instability.
A similar situation holds for $s=0$ solutions, but in this case
the opposing gradients are behind the contact discontinuity
rather than in front.

While the idealized self-similar hydrodynamical model has been very
successful in describing general aspects of SNR evolution, it is
clearly incomplete in some important ways.
For instance, the hydrodynamic model ignores any effects from a
magnetic field, the one-dimensional approximation eliminates the
ability to describe irregularities in the ambient conditions or \RT\
generated instabilities, and particle acceleration at the
collisionless forward and reverse shocks is neglected.
Here we focus on how efficient cosmic-ray production influences 
R-T instabilities by setting $\gameff < 5/3$.
We find that lowering $\gameff$ causes the blast wave to 
expand slightly slower, 
but the radius of the forward shock still follows
the self-similar result $R_1 \propto t^{(n-3)/(n-s)}$. More
importantly, lowering $\gameff$ causes
the width of the region between the forward and reverse shocks to
shrink considerably and alters the internal structure of the
self-similar driven wave.

\begin{figure}[!hbtp]       
\plotone{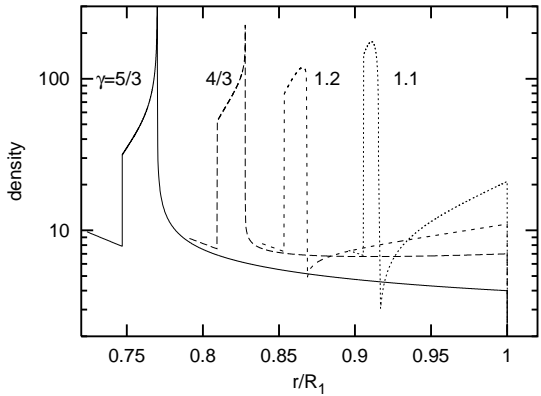}
\figcaption{Radial profile of the gas density in SSDWs with
$n=7$ and $s=2$ for values of $\gameff$ as labeled.  Note the
change in sign of the density gradient behind the forward shock.
Here and in Figure~\ref{fig:onedim}, 
the density is scaled to its value immediately ahead of
the forward shock.
\label{fig:den_prof}}
\end{figure}

We list some of the parameters describing SSDWs
for a range of $\gameff$ in Table~\ref{tbl-1} and, for $\gamma=5/3$, 
these values are identical to those in Table 1 of
\citet{rog82a}.  These parameters, as
well as the radial profiles shown in Figure \ref{fig:den_prof}, were
generated following the procedure outlined in \citet{rog82a}.  Note in
particular that the value of $b$, which contains the $\gameff$
dependence, changes by less than 30\% as the compression ratio is
varied from 4 to 21.  Thus the velocity of the blast wave is not
strongly affected by the value of the compression ratio.  
In contrast, between
$\CR=4$ and $21$, the width of the interaction region shrinks by a
factor of 4 in the case of $n=7$ and $s=0$.

The radial profiles of SSDWs for different values
of $\gameff$ are shown in Figure \ref{fig:den_prof}.  While the
increase in the compression ratio and decrease in the width of the
interaction region are clearly evident, there is
also a change in the density gradient of the shocked
CSM.  As gas moves from the forward shock to the contact discontinuity
it compresses/expands adiabatically to match the relatively constant
pressure of the interaction region.  A lower $\gameff$ leads to a slower
drop in pressure due to radial expansion, so the gas does not need to
be compressed as much as with a higher $\gameff$.  The result is a more
positive density gradient in the shocked CSM for a smaller
$\gameff$.  From Figure \ref{fig:den_prof} we see that this can even
lead to a change in the sign of the density gradient, which may in turn
affect the stability properties of the waves.  Nonetheless, we
expect the low $\gameff$ driven wave to
remain unstable, as there is still a shell of dense shocked ejecta
being decelerated by lower-density shocked CSM.

\section{HYDRODYNAMIC SIMULATIONS}\label{sec:hydro}

To examine the effects of different compression ratios on the
convective stability of driven waves, we use the Virginia Hydrodynamics
(VH-1) time-dependent hydrodynamic numerical code in one, two, and
three dimensions.  All of the simulations were
computed in spherical geometry on a numerical grid with 500 zones in
each dimension.  This resolution is comparable to the highest
resolution simulations of \citet{cbe92}.  Based on their results, we
expect that a higher resolution simulation would show more small scale
mixing, but the large-scale dynamics of the problem would not change.
The angular extent of the simulation domain was chosen to produce roughly
square numerical zones, i.e., $R \Delta\theta \approx \Delta R$.
Periodic boundary conditions were applied in both the $\theta$ and
$\phi$ directions for the multidimensional simulations.  The radial
boundary conditions were set to match the relevant power laws.  The
numerical grid was expanded to follow the forward blast wave so the
evolution could be tracked for many expansion times.

We ran one-dimensional (1D) and two-dimensional (2D) 
simulations with the four combinations
of $s=0,2$ and $n=7,12$.  For each $(n,s)$ pair we ran four
simulations with $\gameff = 5/3, 4/3, 1.2$, and 1.1.  In the following
discussions we will focus on the set of simulations with $n=7$ and
$s=2$ as a specific example, but the results are qualitatively the
same for all of these parameter choices.

\subsection{1-D SIMULATIONS}

We ran 1D simulations for all the cases listed in Table 1
as a check on our ability to numerically recreate the analytic,
self-similar solutions.  The 1D hydrodynamic simulations
were initialized with appropriate density power laws, and allowed to
evolve until a self-similar state was reached.  The gas pressure in
the ejecta and CSM was kept sufficiently low to ensure that the Mach
numbers of the forward and reverse shocks were always greater than
100.

\begin{figure}[!hbtp]   
\plotone{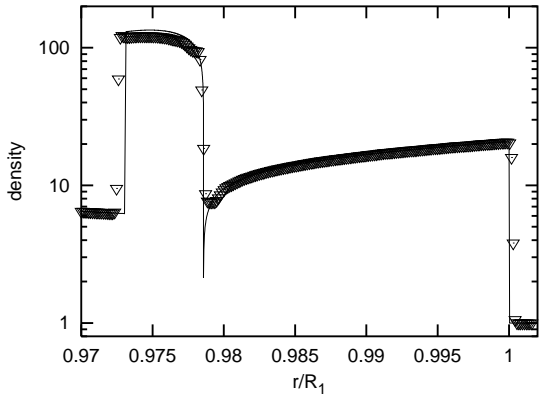}
\figcaption{
The density profile of a driven wave for the relatively extreme case
with $n=12$, $s=0$, and $\gameff=1.1$.  The solid line is the
self-similar analytic solution and the triangles are from the
1D hydrodynamic simulation.
\label{fig:onedim}}
\end{figure}

In Figure~\ref{fig:onedim} we show the results of our most extreme case
with $n=12$ and $s=0$ and $\gamma=1.1$ overlayed on the semi-analytic
solution obtained by numerically integrating an ordinary differential
equation \citep{rog82a}.  This particular SSDW is very thin and possesses
steep gradients, making it a difficult test case for the hydrodynamic
code.  Nonetheless, our simulation does a reasonably good job of matching
the analytic solution; note in particular the
relative sharpness of the shock fronts and contact discontinuity.  
In the less extreme cases,
the match between the simulation and the analytic solution is better
than shown in Figure~\ref{fig:onedim}.
However, the very sharp density peak at the contact discontinuity in
the $s=2$ solutions, and the density trough in the $s=0$ solutions,
are typically smoothed out over a few zones by this hydrodynamic
method.

\subsection{2-D SIMULATIONS}

The multi-dimensional simulations were initiated with the analytic
SSDW solutions normalized to $R_1=1$, $V_1=1$, and
a preshock density of unity.  The simulation thus begins at an
initial time of $t=(n-3)/(n-s)$ in these normalized units.  The
initial solution is mapped
onto a spherical grid with a radial span roughly
twice the width of the interaction region.  The convective
instability was seeded by adding acoustic noise to the shocked
interaction region.  These simulations were evolved long enough for
the convective instability to reach saturation, such that the mixing
region is in a quasi-self-similar state with the growth of R-T fingers
matched by their destruction due to shearing and advection.  This
typically required 
$\sim 6$--8 doubling times of the blast wave radius,
and beyond this time, the results are qualitatively independent of the seed
noise.

\begin{figure}[!hbtp]     
\epsscale{0.75}
\plotone{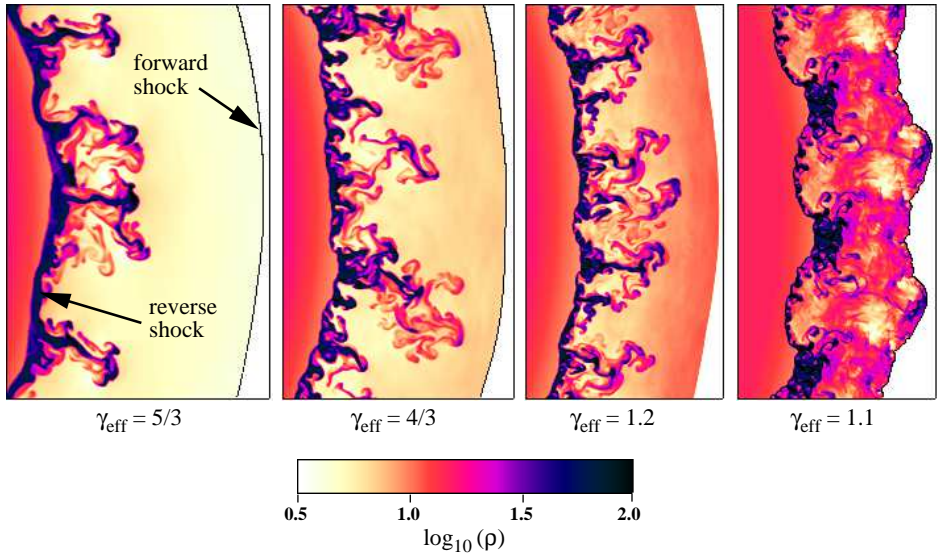} 
\figcaption{Convective instabilities in driven
waves with $n=7$ and $s=2$ and values of $\gameff$ as marked. 
The color scale represents gas density, scaled to the density of
the CSM immediately ahead of the forward shock.  The thin
shell of dense, shocked
ejecta is deformed into narrow fingers, characteristic of the R-T
instability.
\label{fig:density}}
\end{figure}

In all of the cases listed in Table \ref{tbl-1}, the layer of shocked
ejecta was found to be unstable with an evolution similar to that seen
in previous studies \citep[\iec][]{cbe92,jun96}.  As an example, we
show the density for driven waves with $n=7$, $s=2$, and different
values of $\gameff$ in Figure \ref{fig:density}.  The results for
other combinations of $n$ and $s$ are similar.  These images are taken
at the end of the simulations, once the instability has reached
saturation.  In addition to shrinking the interaction region, a lower
value of $\gameff$ produces a narrower, denser shell of shocked ejecta
and sharper, denser R-T fingers.  Aside from the narrower interaction
region, these results for small $\gameff$ are similar to those found
by \cite{cbl95} for driven waves with radiative cooling in the shocked
ejecta.  In this latter case the ejecta behaved as an ideal gas with
$\gamma \approx 1$, while the CSM was modeled as an ideal gas with
$\gamma = 5/3$.

\onecolumn
\begin{table}[!hbtp]
\begin{center}
\caption{
Numerically derived parameters describing spherically symmetric driven waves.
The subscript $1$ ($2$) indicates values immediately behind the forward (reverse) shock.
\vskip12pt
\label{tbl-1}}
\begin{tabular}{crrcccrrrrrr}
\tableline\tableline
$s$ & $n$ & $\CR$ & $\gamma$ & $R_2/R_1$ & $b$ & 
$\rho_2/\rho_1$ & $p_2/p_1$ & $u_2/u_1$ \\
\tableline
0 & 7 &  4 & 1.67 & 0.792 & 1.212 & 1.34 & 0.47 & 1.253\\
0 & 7 &  7 & 1.33 & 0.865 & 1.125 & 1.21 & 0.51 & 1.118\\
0 & 7 & 11 & 1.20 & 0.907 & 1.081 & 1.14 & 0.53 & 1.066\\
0 & 7 & 21 & 1.10 & 0.947 & 1.044 & 1.08 & 0.54 & 1.030\\
\tableline
2 & 7 &  4 & 1.67 & 0.747 & 0.996 &  7.8 & 0.27 & 1.058\\
2 & 7 &  7 & 1.33 & 0.810 & 0.897 &  7.5 & 0.31 & 0.978\\
2 & 7 & 11 & 1.20 & 0.853 & 0.840 &  7.2 & 0.33 & 0.960\\
2 & 7 & 21 & 1.10 & 0.906 & 0.782 &  6.8 & 0.35 & 0.962\\
\tableline
0 & 12 &  4 & 1.67 & 0.869 & 0.977 & 7.2 & 0.60 & 1.255\\
0 & 12 &  7 & 1.33 & 0.922 & 0.925 & 6.7 & 0.63 & 1.128\\
0 & 12 & 11 & 1.20 & 0.950 & 0.902 & 6.4 & 0.64 & 1.076\\
0 & 12 & 21 & 1.10 & 0.973 & 0.882 & 6.2 & 0.66 & 1.038\\
\tableline
2 & 12 &  4 & 1.67 & 0.805 & 0.884 &  46 & 0.37 & 1.104\\
2 & 12 &  7 & 1.33 & 0.866 & 0.817 &  43 & 0.40 & 1.026\\
2 & 12 & 11 & 1.20 & 0.903 & 0.780 &  41 & 0.41 & 1.003\\
2 & 12 & 21 & 1.10 & 0.942 & 0.745 &  39 & 0.43 & 0.994\\
\tableline
\end{tabular}
\end{center}
\label{numbers}
\end{table}

\twocolumn

\begin{figure}[!hbtp]        
\epsscale{1.0}
\plotone{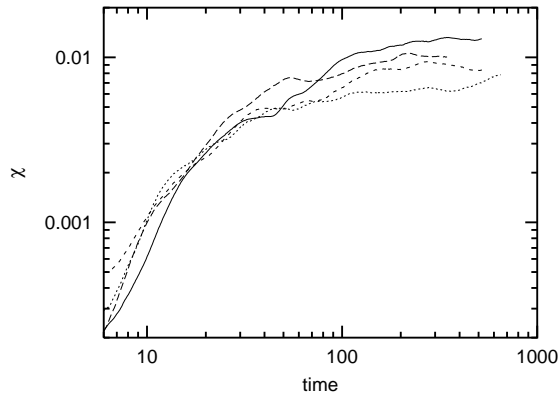}
\figcaption{Growth of the convective instability as measured 
by the normalized turbulent energy density, $\chi$, in the interaction region.
These curves are from 2D simulations with
$n=7$, $s=2$, and $\gameff =$ 5/3 (solid), 4/3 (long dash),
1.2 (short dash), and 1.1 (dotted).
\label{fig:growthcurve}}
\end{figure}

We can obtain a rough estimate of the instability growth rate by
following the turbulent energy density in the intershock region.
Rather than attempt to subtract off the bulk radial velocity to find
the contribution to the turbulent velocity, and to avoid confusion in
comparing 2D and 3D simulations, we consider only a single angular
component of the turbulent motion.  Furthermore, to remove the effects
of radial expansion, deceleration, and shock compression, we normalize
this turbulent energy to the bulk kinetic energy density associated
with the shock front.  We thus define the ratio
\begin{equation}
\chi = \frac{\int \rho u_\theta^2 d\tau}{\rho_1V_1^2 \int d\tau}
\ ,
\end{equation}
where the integration is only over the volume of shocked gas.  The 
evolution of this
turbulent energy parameter is shown in Figure \ref{fig:growthcurve},
from which one can see that while the saturated levels of
turbulent energy density differ by about a factor of two, 
the growth rates are not noticeably
affected by changes in $\gameff$, despite the fact that the radial
profile from the spherical solution has changed dramatically (e.g.,
the density gradient has changed sign).

\begin{figure}[!hbtp]               
\plotone{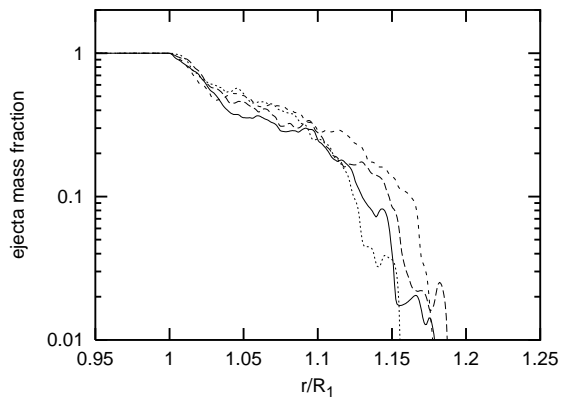}
\figcaption{The width of the mixing region, as illustrated 
with the mass fraction of ejecta, is relatively unaffected
by changes in $\gameff$.  For this comparison we have 
normalized radii in each model
to the reverse shock rather than the forward shock, so we 
can directly compare the radial extent of the mixing region.
The line styles are as in Figure~\ref{fig:growthcurve}.
\label{fig:fingers}}
\end{figure}

The saturation of the instability, at least in terms of the width of
the mixing region, is also relatively unaffected by changes in
$\gameff$.  In Figure \ref{fig:fingers} we plot an angle-averaged
radial profile of the ejecta mass fraction.  In a
spherically-symmetric model this would be a step function from unity
dropping to zero at the contact discontinuity.  To provide an easy
comparison, we have normalized the radii such that the reverse shock
is located at the same radius for all models.  From this we can see
that the length of the R-T fingers is comparable for all four
values of $\gameff$.

\begin{figure}[!hbtp]        
\plotone{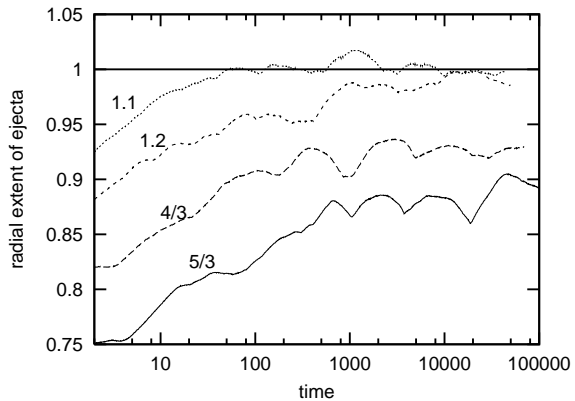} \figcaption{The maximum radial extent 
(measured relative to the average radius of the outer shock) of the R-T
fingers of dense shocked ejecta is shown as a function of time for
$n=7$ and $s=2$.  For the smallest value of $\gameff$, the fingers
reach all the way to the shock front.  For the other values of
$\gameff$ the R-T fingers reach a maximum length after $t\sim 1000$.
\label{fig:extent}}
\end{figure}

While the instability itself is insensitive to changes in the
compression ratio, the overall width of the interaction region
decreases with increasing compression ratio.  This has the important
effect that, for small $\gameff$, the R-T fingers can reach as far as
the forward shock front.  To measure the maximum extent of the R-T
fingers, we identify an average mass fraction of
0.001 as the leading edge of the mixing region.
Since the radial gradient
of the mass fraction is quite steep at these small values, the
exact choice of a cutoff value does not affect the results.  The time
dependence of the maximum radial extent (relative to the forward
shock) of the R-T fingers is illustrated in Figure \ref{fig:extent}.
In the run with the highest compression ratio ($\gameff=1.1$), the R-T
fingers quickly reach and pass the average shock radius.
We note that when the curves in Figure \ref{fig:growthcurve} become
approximately flat, the system is in a near self-similar state and the
ratio of the forward shock radius to the contact discontinuity radius
remains approximately constant. If we allowed the system to evolve to
the point where the reverse shock entered the plateau region in the
ejecta density (which nominally happens when the mass of swept up CSM
is comparable to the ejecta mass), the ratio of the forward shock
radius to the contact discontinuity radius would start to increase and
the R-T fingers would drop behind the forward shock.

Once the fingers of dense, shocked ejecta reach the forward shock,
they can begin to distort the outer blast wave by pushing small regions
out ahead of the average shock radius.  The resulting bumps in the
outer shock can be seen in Figure \ref{fig:density} for the simulation
with $\gameff = 1.1$.  The shock front is nearly spherical for large
values of $\gameff$, while values close to unity produce significant
deviations from spherical symmetry.

We note that even in the most dramatic case of $n=12$, $s=0$, and $\gameff
= 1.1$, the deviation from a spherical outer shock 
remains relatively small; the penetrating fingers
are not able to dramatically alter the forward shock.  This can
be seen in Figure \ref{fig:extent}, where the maximum radial extent of
the fingers appears capped at $\sim R_1$. 
Since the instability is relatively unaffected by the value of
$\gameff$, we expect Figure \ref{fig:extent} to show a series of
parallel curves as the fingers grow until saturation occurs (at
several 100 times the initial simulation time, although this
depends on the magnitude of the initial perturbation).  
If the fingers reach the forward shock
before saturation, as in the $\gameff = 1.1$ case, they cannot push
out much beyond the shock and their growth stalls.  This limited
ability to affect the shock front is due to the strong shearing flow
created when the shock front is distorted.  If a clump of dense ejecta
has sufficient radial momentum to push out the forward shock in a
local protrusion, the deformed shock front generates a substantial
tangential post shock flow around the ejecta clump.  In all of our
simulations, this shear flow quickly disrupts the clump through the
Kelvin-Helmholtz instability, with the remnants of the clump quickly
advected back into the interaction region.  Thus we never see a local
protrusion of the shock front stick out more than a few percent of the
blast wave radius.

\subsection{3-D SIMULATIONS}

We repeated two simulations in 3D to check
that these results are not dramatically affected by the
assumption of axisymmetry, using $n=12$ and $s=0$ for 
$\gameff = 5/3$ and 1.1.  In comparing the 2D
and 3D simulations, we stress that the
numerical code remained virtually identical.  The only 
changes involved using a 3D grid and repeating the hydrodynamic
updates in a third direction.  

\begin{figure}[!hbtp]            
\epsscale{0.78}\plotone{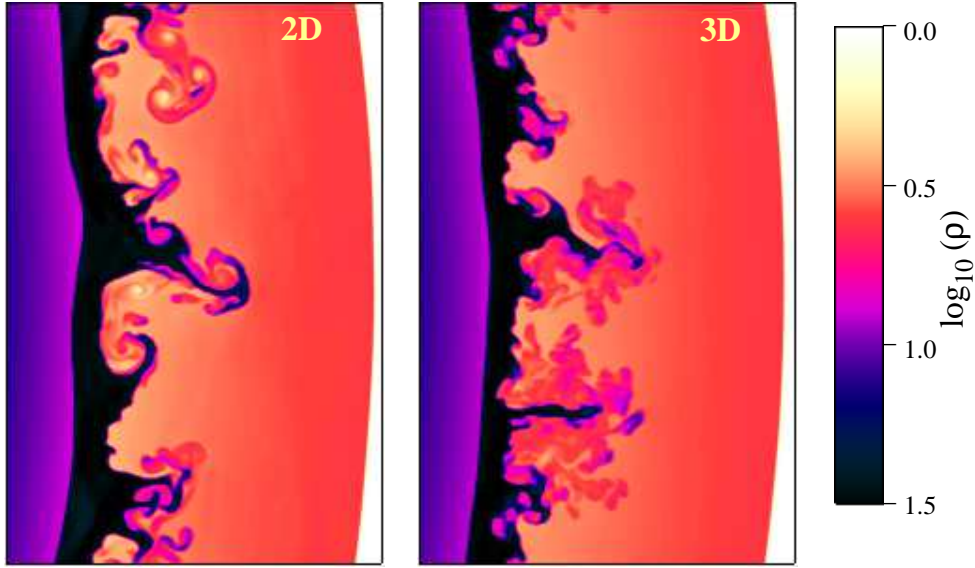} 
\figcaption{A comparison of the shock structure in
a 2D axisymmetric simulation (left), with a planar slice
from a 3D simulation (right).  The shading depicts 
gas density, scaled to the density of the preshock gas.
Black represents the high density of the shocked
ejecta and white the low density of the unshocked CSM.  
These runs used $n=12$, $s=0$, and $\gameff=5/3$.
\label{fig:threed53}} 
\end{figure}

For the more familiar case of $\gameff = 5/3$ our results are
quite similar to previous 3D simulations of driven
waves \citep[\egc][]{bij96}, showing only minor differences between
2D and 3D simulations.  This similarity is exhibited in Figure \ref{fig:threed53}.
The only significant difference visible in this Figure is the increased
amount of small-scale structure in the 3D simulation, although this
difference does not appear to affect any of the global properties of
the driven wave.  A quantitative comparison of
these two simulations finds that the
growth and saturation of the turbulent energy is virtually 
identical, the radial extent of the fingers is comparable (the
3D fingers reached slightly further than the 2D fingers), the
forward shock remains spherical and the width of the interaction
region grows slightly in both cases as the instability approaches
saturation.  

\begin{figure}[!hbtp]            
\epsscale{0.68}\plotone{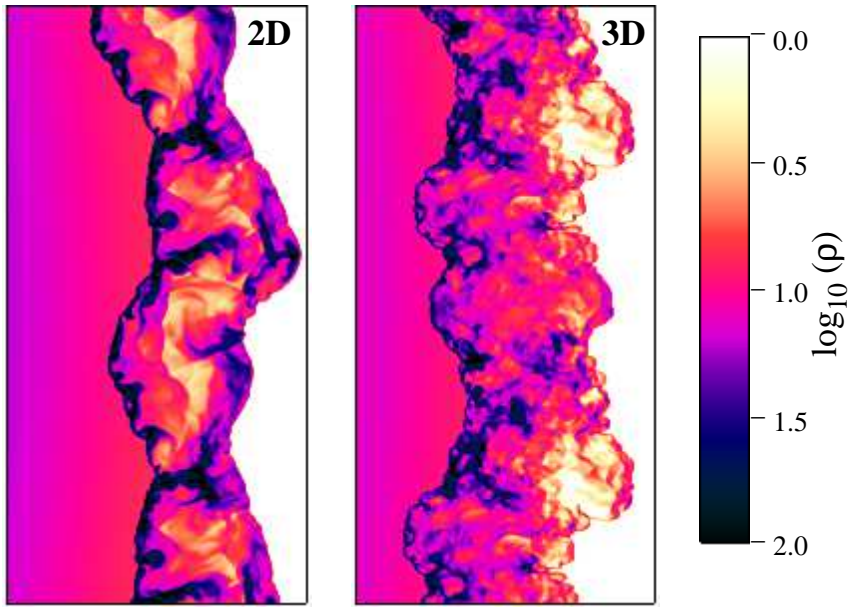} \figcaption{A comparison of the shock structure in
a 2D axisymmetric simulation (left), with a planar slice
from a 3D simulation (right).  These runs used $n=12$,
$s=0$, and $\gameff=1.1$. 
\label{fig:threed}} 
\end{figure}

The situation is somewhat more complicated for 
the case of high shock compression, as shown in Figure \ref{fig:threed}.
Here one sees much more small-scale structure in the 3D simulation,
both within the interaction region and in the forward and reverse 
shocks.  This more dramatic difference between 2D and 3D in
the high-compression case is consistent with the fact that the
R-T fingers do not reach the shock front when the shock compression is
low, but they do reach - and perturb - the forward shock when
the compression is large.  For both values of $\gameff$ we see more
small-scale structure in the R-T fingers in 3D, but in the high-compression
case this small-scale structure can modify the forward shock to produce
many small wavelength, but large amplitude, perturbations.

A consequence of the more oblique shocks created by the deformation
of the forward shock in the 3D
simulation is an overall drop in the average compression ratio
and a corresponding increase in the width of the interaction region.
Despite these differences, our statistical measures of 
the convective instability remain relatively unaffected
by the dimensionality of the simulation.
Furthermore, our primary
conclusion from the 2D simulations still holds in 3D; the R-T fingers
are able to deform the forward shock when the compression ratio is high.

\onecolumn

\twocolumn

\section{DISCUSSION}\label{sec:disc}

If young supernova remnant shocks accelerate cosmic rays with high
efficiency, the acceleration process can cause these high Mach number
shocks to have compression ratios considerably greater than four.  We
have investigated the effects of this high compression on the
hydrodynamic stability of young remnants with a simple and direct
method; we perform hydrodynamic simulations with the effective
adiabatic index, $\gameff$, set to values less than $5/3$, yielding
compression ratios, $\CR \simeq (\gameff +1)/(\gameff - 1) > 4$.
While this fluid approach is clearly an approximation to the real
situation where the collisionless shocks accelerate particles to
relativistic energies, we believe it accounts for the major effects of
efficient particle acceleration on the overall dynamics and, most
importantly, on the \RT\ instabilities that develop in the interaction
region between the forward and reverse shocks.

Somewhat surprisingly, we find that a large compression ratio has
little effect on the {\it growth} of R-T instabilities. However, since
a large $\sigma$ dramatically shrinks the {\it width} of the
interaction region between the forward and reverse shocks, several
important effects emerge from of our high-compression simulations:

(1) In contrast to shocks with $\CR \sim 4$, R-T fingers reach closer
to the forward shock front, and if $\CR$ is large enough, ejecta
material can be found at, and even slightly ahead of the average shock
radius.  This can occur fairly quickly after the explosion (depending
on how quickly a reverse shock forms and the magnitude of any 
inhomogeneities seeding this instability)
and the fingers should stay near the forward shock front
throughout the time the reverse shock is in the power law portion of
the ejecta density profile.  In general this should apply to
SNRs from an age of only months up to several 1000 years.

(2) The forward shock is perturbed on short wavelengths and with
relatively small amplitudes.  In addition to slightly altering the
spherical shape of the shock front, this will lead to some small
spread in the postshock temperatures.  Note also that the overall
temperature of the shocked gas will be substantially lower if particle
acceleration is efficient and compression ratios are large than if
little acceleration occurs \citep[\egc][]{EBB2000}.  This effect
provides a coupling between the shock morphology and thermal X-ray
emission \citep{Decourchelle00,Hughes00b}.

(3) If the mixing region reaches the forward shock, the morphology of
the magnetic field as seen through radio synchrotron polarization
observations may be affected. \citet{bij96} followed the evolution of
the ambient magnetic field in driven wave simulations, looking for an
explanation for the origin of observed polarization in young SNRs in
the elongation of the field by R-T instabilities. Our work offers a
ready explanation for why this polarization can extend all the way to
the forward shock.

While we do not include magnetic fields in the hydrodynamic
simulations we perform here, we note that any ambient magnetic fields 
will also be compressed at the shock front.
In addition, the increased turbulence associated with
the high compression driven waves may be expected to further 
amplify the magnetic fields and show stronger
radio emission and possibly more intense TeV emission from \IC\ and/or
\pion\ than the more quiescent driven waves found for $\gameff = 5/3$.
It can be expected that the
effects we see here from increased compression will add to those
reported by \citet{JJ99}.

The SNR \SNRsmc\ in the Small Magellanic Cloud may represent a young remnant
for which this model of a high compression driven wave is applicable.  
Recently, \citet{Hughes00b} have determined postshock electron
temperatures in SNR \SNRsmc\ using {\it
Chandra X-Ray Observatory} observations.  Using the measured forward
shock speed of $\Vfs \sim 6000$ \kmps, they find that the electron
temperature of 0.4--1 keV is at least 2.5 times lower than can be
explained with standard (i.e., test-particle) shock heating even if
only Coulomb electron heating occurred.  They
conclude that the forward shock is placing at least 50\% of the
shock kinetic energy flux in cosmic ray ions.

\begin{figure}[!hbtp]            
\epsscale{1.0}\plotone{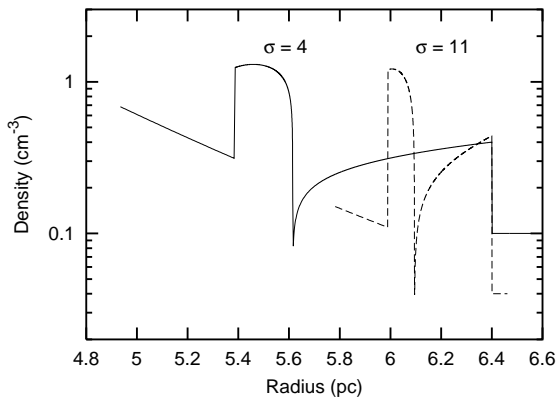} 
\figcaption{The radial profile of density ($\rho/m_p$) for a model of
\SNRsmc\ with efficient particle acceleration assuming $\gameff=1.2$ (dashed curve), 
and one with test-particle acceleration assuming $\gameff=5/3$(solid curve), both at an
age of $\tSNR \simeq 700$ yrs.
\label{fig:SMCsnr} } 
\end{figure}

If one applies a SSDW model to the observed radius and expansion 
velocity of \SNRsmc\ \citep[e.g.,][]{Gaetz2000,Hughes00b}
assuming $n=9$ and $s=0$,
one finds an age of 700 years, independent of the compression ratio.
Choosing reasonable parameters of $\EnSN= 10^{51}$ erg and $\Mej=1 \Msun$,
one finds an ambient density of $\rho_a \approx 0.1\, m_p$ \pcc
for $\sigma = 4$.  If instead we 
assume a large compression ratio of $\sigma = 11$, the derived ambient density
is lower by only a factor of 2.3.  Despite the relative similarity in
fitting the observed SNR parameters, these models, shown in Figure \ref{fig:SMCsnr}
are dramatically different in terms of the 
position of the reverse shock and the postshock temperature.
If the shock compression is high, the reverse shock will be significantly
closer to the forward shock, as shown in Figure \ref{fig:SMCsnr}.  This is
not consistent with X-ray observations which show the presence of shocked
ejecta at a radius of only $\sim 70\%$ of the forward shock \citep{Gaetz2000}.
However, if the ejecta mass is relatively small, this SNR may be evolving
away from the SSDW phase, with the reverse shock now propagating in toward
the center of the SNR.

With these values and one additional parameter, the ambient magnetic
field strength $B_0$, we can estimate the particle acceleration using
the model of \citet{BEapj99}.
We find that with $B_0 \simeq 3$ \muG, the forward shock obtains
$\sigma \simeq 11$ (consistent with $\gameff=1.2$) with a shocked
proton temperature of $T_{p2} \simeq 4\xx{7}$ K, i.e., 10 times lower
than obtained in a test-particle shock with the same parameters (for
the acceleration calculation we assume a constant $\Vfs = 6000$ \kmps\
for $\sim 700$ yr).  The low postshock electron temperature deduced
from X-ray observations is now easily explained if the electron
temperature is $\sim 1/3$ $T_{p2}$.\footnote{If we had used a lower
$\gameff$, we could have obtained a consistent fit with a smaller
$T_{p2}$.}
Furthermore, the forward shock accelerates
electrons to $\sim 70$ TeV in $\sim 700$ yr, consistent with radio
\syn\ emission.  At the current age, $\MSZ \simeq 3700$ and $\MAZ \simeq
150$, and the acceleration is extremely efficient, i.e., more than
80\% of the kinetic ram energy flux is placed in relativistic ions.
While this model is not unique, it does show that efficient particle
acceleration is consistent with reasonable supernova and ambient
medium parameters, as well as with deduced values of radius, speed,
age, and electron temperature.

While there are clearly differences between the actual situation in
SNRs where some fraction of swept-up material is shock accelerated to
relativistic energies, and simply lowering $\gameff$, we believe our
results are qualitatively correct.  The important differences include
the fact that lowering $\gameff$ implies that the effects of
relativistic particle pressure and particle loss occur everywhere
rather than just in the shocked gas and in the precursor regions in
front of the forward and reverse shocks. We note that if the finite
size of the shock is limiting the maximum particle energy, there will
be a significant fraction of the total pressure in high energy
particles with upstream diffusion lengths on the order of $1/10$ of
the shock radius.
This difference in pressure distribution will have some effect on the
shock evolution since, instead of a uniform $\gameff$, the actual
remnant has a gas with a soft equation of state pushing one with a
harder equation of state. While we have not tested this difference in
detail, we expect the effects we found will actually be enhanced if
$\gameff$ varies spatially.

Another difference comes about because we use eq.~(\ref{eq:sigma})
with a constant $\gameff$ to determine $\sigma$ instead of
eqs.~(\ref{eq:machsonic}). With
(\ref{eq:sigma}), there is little variation in $\sigma$ as the SNR
evolves as long as $\MSZ \gg 1$, while eqs.~(\ref{eq:machsonic}) 
can give a much larger variation depending on the parameters.
\footnote{
We note that it is not always the case that
lowering the Mach numbers causes $\CR$ to
decrease. Equations~(\ref{eq:machsonic}) apply
when injection into the shock acceleration mechanism is efficient. If
injection is weak enough, high Mach number, test-particle solutions
with $\CR \simeq 4$ can result \citep[see][for a full
discussion]{BEapj99}.
}
In any case, since in all probability $\sigma \gg 4$ during the time
we consider before the reverse shock enters the ejecta plateau region,
any differences resulting from $\sigma$ varying with time are likely
to be small and go in the direction of increasing the effects we
report.

Finally, since the shrinking of the interaction region between the
forward and reverse shocks depends totally on the pressure in that
region, effects other than particle acceleration which influence the
pressure may be important. The most likely effect which we have neglected
comes from the compression of the magnetic field, $B$. Since the magnetic
pressure scales as $B^2$, increasing the  compression ratio could
produce magnetic pressures large enough to prevent the interaction
region from becoming narrow enough to allow the R-T fingers to reach
the forward shock. This, of course, will depend on the Alfv\'en Mach
number and the angle the upstream field makes with the shock.
The pressure effects of the magnetic field will be offset somewhat by
the fact that, for a given compression ratio, the shocked thermal
pressure is considerably less in a shock undergoing efficient
acceleration compared to one with a low $\gameff$ and no acceleration.

\acknowledgements We thank A. Decourchelle, J. P. Chi\`eze, and
S. Reynolds for helpful discussions.
The numerical simulations reported here were performed
at the North Carolina Supercomputing Center using 100 processors of
an IBM SP2.  We thank NCSC and IBM for their generous support of computing 
resources.
Support for this work was provided by NASA under grant NAG-7153.

\onecolumn

\end{document}